# Two-Dimensional h-BN and MoS$_2$ as Diffusion Barriers for Ultra-Scaled Copper Interconnects

Chun-Li Lo[1], Massimo Catalano[2,3], Kirby K. H. Smithe[4], Luhua Wang[2], Shengjiao Zhang[1], Eric Pop[4,5,6], Moon J. Kim[2], and Zhihong Chen[1*]

[1] School of Electrical and Computer Engineering and Birck Nanotechnology Center, Purdue University, West Lafayette, IN 47907, USA

[2] Materials Science and Engineering Department, The University of Texas at Dallas, 800 West Campbell Road, Richardson, TX 75080, USA

[3] Institute for Microelectronics and Microsystems, National Council for Research (IMM-CNR), Via Monteroni, ed. A3, 73100 Lecce, Italy

[4] Department of Electrical Engineering, Stanford University

[5] Department of Materials Science and Engineering, Stanford University

[6] Precourt Institute for Energy, Stanford University, Stanford, CA 94305, USA

## Abstract

Copper interconnects in modern integrated circuits require ultra-thin barriers to prevent intermixing of Cu with surrounding dielectric materials. Conventional barriers rely on metals like TaN, however their finite thickness reduces the cross-sectional area and significantly increases the resistivity of nanoscale interconnects. In this study, a new class of two-dimensional (2D) Cu diffusion barriers, hexagonal boron nitride (h-BN) and molybdenum disulfide (MoS$_2$), is demonstrated for the first time. Using time-dependent dielectric breakdown measurements and scanning transmission electron microscopy coupled with energy dispersive X-ray spectroscopy and electron energy loss spectroscopy, these 2D materials are shown to be promising barrier solutions for ultra-scaled interconnect technology. The predicted lifetime of devices with directly deposited 2D barriers can achieve three orders of magnitude improvement compared to control devices without barriers.

Copper has been used as the most common interconnect material because of its superior conductivity. However, Cu diffusion into the dielectric between two interconnects can cause shorting and create chip failures, while diffusion to transistors can introduce deep-level traps to Si[1] and affect overall transistor performance, as illustrated in Figure 1a. To prevent these undesired effects, some conventional (Ta/TaN, or TiN based) and emerging materials (Ru/Ti, CuMn, etc.)[2,3], have been used or proposed as diffusion barriers by isolating Cu from surrounding intra- and inter-layer dielectrics. However, these barrier materials are all at least one order of magnitude more resistive than the Cu itself[4]. Thus, to maximize the Cu volume in its damascene trench for lower line resistance, the thickness of the barrier material must be reduced as much as possible. Conversely, it has been found that conventional barrier materials lose their capability of blocking Cu diffusion when their thicknesses are scaled below ~3 nm, as illustrated in Figure 1b. According to the International Technology Roadmap for Semiconductors (ITRS)[5], ultra-thin diffusion barrier materials with thickness close to 1 nm are highly demanded in the near future.

Graphene has recently been demonstrated to have superior capability of blocking Cu diffusion despite its atomic thickness[6-8] and has been shown to enhance the electrical and thermal conductivity of Cu[9]. In the meantime, a group of other two-dimensional (2D) layered materials exists, whose properties are complementary yet distinct from those of graphene. Theoretical calculations predict high-energy barriers in some of these materials to prevent molecule diffusion[10,11]. In the development of conventional diffusion barrier materials, various material types including both metals and insulators have been investigated, judged by the interface requirements of different applications[12,13]. While it is still a rather unexplored field with many unknowns such as Cu wetting and adhesion, interface scattering, and CMOS compatibility, it is important to evaluate the potential of these atomically thin 2D materials as ultra-thin barriers and make thorough comparisons.

In this work, the diffusion barrier properties of 2D hexagonal boron nitride (h-BN) and molybdenum disulfide ($MoS_2$) are investigated by time-dependent dielectric breakdown (TDDB) measurements. h-BN is an atomically thin 2D insulator (band gap ~6 eV)[14] and $MoS_2$ is an atomically thin semiconductor (band gap ~2 eV)[15]. We observe that the lifetime of intra- and inter-layer dielectrics can be significantly extended with the presence of the tested 2D barriers. In addition, using scanning transmission electron microscopy (STEM), energy dispersive X-ray spectroscopy (EDS), and electron energy loss spectroscopy (EELS), we confirm that the examined ultra-thin 2D barriers can efficiently mitigate Cu ion diffusion. Our results provide strong evidence for promising alternative barriers using both insulating and semiconducting 2D materials. In general, the demonstrated 2D barriers can also be useful in other applications where preventing undesired mass transport or corrosion is important.

## Results and Discussions

**Device structure and material preparation.** To perform electrical measurements, a metal-oxide-semiconductor (MOS) capacitor structure was fabricated, as shown in Figure 1c. Details of the material preparation and device fabrication/structure are described in the "Methods" section. Devices with 2D barrier layers inserted between Cu and $SiO_2$ were evaluated for the diffusion barrier properties, while devices without any 2D barriers were prepared as control samples. Three types of barrier samples were compared: 1) h-BN grown by chemical vapor deposition (CVD) on a Cu foil was transferred onto a 20 nm $SiO_2$/Si substrate to form a three to four-layer (3-4L) h-BN barrier; 2) Single-layer (1L) $MoS_2$ with some small two-layer (2L) regions was directly grown on a 30 nm $SiO_2$/Si substrate by CVD[16] at 850 °C; 3) 1L $MoS_2$ from the same CVD growth was transferred to a 20 nm $SiO_2$/Si substrate for a direct process comparison that will be discussed. Note that, it may be possible in the future to lower the growth temperature to be back-end-of-line (BEOL) compatible (e.g. by atomic layer deposition), which is not the focus of this work. The STEM cross-section image in Figure 2a reveals that there may be small thickness variations in our h-BN sample, ranging between 3-4 layers. Optical images suggest that the CVD-grown MoS$_2$ films consist of continuous and discontinuous regions, marked in Figure 2b. Electrical measurements that will be discussed were conducted on devices fabricated in selected continuous regions. 1L MoS$_2$ was verified by Raman spectroscopy, with characteristic E' peak at 384.5 cm$^{-1}$ and A$_1$ at 405 cm$^{-1}$, respectively, as shown in Figure 2c. Note that the Raman peak separation ($\triangle f \sim 20$ cm$^{-1}$) is slightly higher than that of exfoliated 1L MoS$_2$ due to tensile strain in as-grown CVD films[17]. In addition, although 1L MoS$_2$ was dominant, characteristic Raman peaks associated with 2L MoS$_2$ (Figure 2c) were also found occasionally (~10% coverage).

**TDDB measurements and lifetime prediction.** To rule out the variations generated by defects and grain boundaries in CVD-grown 2D films, or other imperfections from the not yet optimized CVD recipes, TDDB was adopted to evaluate the diffusion barrier properties of these 2D materials since it provides a statistical approach for a fair analysis. In addition, TDDB has been widely accepted as a test vehicle for assessment of Cu interconnect reliability[18-25]. In our TDDB setup, a positive constant electric field (E-field) was applied at room temperature to the top Cu electrode of the device-under-test, with the bottom p$^{++}$ Si being grounded, as shown in Figure 1c. If the positive E-field drives Cu ions into SiO$_2$, these ions can accumulate and form a conductive path in the dielectric and/or assist in Poole-Frenkel tunneling[21], which leads to device breakdown. Time-to-breakdown (t$_{BD}$) of each device was recorded when the device broke down and the leakage current density reached $1.3 \times 10^{-2}$ μA/μm$^2$ (equivalent to 100 μA from a circular metal pad with 100 μm diameter). Once t$_{BD}$ of more than 10 devices (more than 15 in most cases) was obtained, an evaluation of the dielectric quality that takes the variability into account was finally achieved. If significant Cu ion diffusion is present, t$_{BD}$ will be reduced due to Cu-induced breakdown, as illustrated in the left part of Figure 1c. If the 2D barrier can mitigate the Cu ion diffusion, t$_{BD}$ is expected to be extended due to the lower probability of conduction path formation, which is depicted in the right part of Figure 1c.

Figure 3a shows the current evolution with time for devices with and without h-BN under a stress of 7 MV/cm. We observe that devices without h-BN barriers reached breakdown earlier in general. Moreover, before the breakdown occurred, the currents of the devices without h-BN were generally higher. Defining the breakdown current at $1.3 \times 10^{-2}$ μA/μm$^2$, t$_{BD}$ of different devices can be obtained from Figure 3a. TDDB results of devices with and without h-BN at various E-fields of 6, 7, and 8 MV/cm are compared in Figure 3b. Each data point represents t$_{BD}$ of a single device. At a certain E-field, the device with the shortest/longest t$_{BD}$ was assigned to have the lowest/highest value of the cumulative probability. Therefore, the slope of the fitted line for any given E-field is always positive. With the presence of the h-BN barrier, t$_{BD}$ of devices has clearly increased, indicating the suppression of Cu ion diffusion. The less-steep slope of the 8 MV/cm line of the h-BN devices is attributed to device variations, which occasionally is inevitable for transferred-CVD films. Despite this, the median-time-to-failure ($TTF_{50\%}$) defined at probability of 0.5 is still a fair indication of the average device reliability since it was statistically obtained from a large number of devices. The purpose of performing TDDB measurements at various E-fields is to allow extrapolation of the device lifetime under normal operating conditions (much lower E-fields) by fitting with some analytic models.[18-23] Otherwise, directly conducting TDDB at low E-fields can be extremely time consuming. Among numerous proposed models, E-model,[18] 1/E-model,[19,20] and sqrt-E-model[21,22] are chosen for low field lifetime predictions, as shown in Figure 3c. The equations of these models with only the E-field dependent terms shown can be expressed as:

$$\text{E-model: } \ln(TTF_{50\%}) \sim -\gamma E \tag{1}$$

$$\text{1/E-model: } \ln(TTF_{50\%}) \sim \left( G/E \right) \tag{2}$$

$$\text{sqrt-E-model: } \ln(TTF_{50\%}) \sim -2\beta_S \sqrt{E} \tag{3}$$

where $\gamma$, $G$, and $\beta_S$ are regarded as constants in this study. While various models emphasizing different breakdown mechanisms have been investigated extensively for decades[23,24], it is well understood that they can vary significantly with different materials, processes, and structures[18-21]. Since detailed breakdown mechanisms are not yet explored in these novel diffusion barrier materials, a lot more research is required to develop sufficient understanding and build models that can eventually provide precise predictions in the future. The models adopted in this study include the most conservative one (E-model) and a relatively optimistic one (1/E-model), based on which qualitative comparisons without detailed mechanism analyses have been accomplished. Our results demonstrate a general enhancement of dielectric lifetime regardless of the model used. In Figure 3c, under the normal operating condition, devices with h-BN have ~50 times longer lifetime (from ~$10^5$ to $7.5 \times 10^6$ s) than devices without barriers, based on the prediction of the E-model.

We now turn to the directly-grown $MoS_2$ barriers. Field dependent TDDB measurement results are plotted in Figure 4a. Based on $TTF_{50\%}$ of the $MoS_2$ devices and the control samples [Supplementary Figure S1b] at different E-fields, comparison of the lifetime prediction is provided in Figure 4b. We observe that, with the presence of the $MoS_2$ barrier, the reliability of the dielectric underneath Cu under normal operating conditions is significantly enhanced, from ~$10^5$ to $3.7 \times 10^8$ s, showing more than three orders of magnitude improvement in device lifetime. It is worth noting that, despite the longer $TTF_{50\%}$ of the devices with transferred h-BN at high E-fields, the predicted lifetime of the devices with directly-grown $MoS_2$ is superior at low E-fields. This discrepancy can be attributed to $SiO_2$ quality degradation due to thermal stress during the CVD growth, which is confirmed in Supplementary Figure S1a. The sulfur-thermal annealed $SiO_2/Si$ sample (labeled as "after 850 °C growth") went through the same CVD process but intentionally received no $MoS_2$ growth. During the CVD growth, the high-temperature facilitated decomposition[26-28] of $SiO_2$ and/or thermal stress-induced diffusion of precursor residues into $SiO_2$ can generate defects in the dielectric. As a result, the sulfur-thermal annealed $SiO_2$/Si sample has lower $t_{BD}$ and higher leakage current before the breakdown. This can be minimized once the growth recipe is optimized. Interestingly, at low E-fields, the extrapolated lifetime of both $SiO_2$/Si control substrates are very similar, as shown by the black curves in Figure 3c vs. Figure 4b. This suggests that the aforementioned CVD-induced $SiO_2$ defects do not contribute much to the reduction of $TTF_{50\%}$ at low E-fields. In contrast, $TTF_{50\%}$ degrades more at high E-fields when the energy barrier for Cu ions to overcome to transport through these defect states is lowered by the E-fields. Therefore, at low E-fields, the lifetime of devices with the transferred h-BN and directly-grown $MoS_2$ can still be compared even though they have gone through different processes. To further verify the proposed mechanism, $MoS_2$ is removed from its original growth substrate and transferred onto the same 20 nm $SiO_2$/Si substrate used for the h-BN samples. As shown in Figure 4c, $TTF_{50\%}$ at high E-fields is higher than that of the directly-grown $MoS_2$ and rather close to the h-BN samples shown in Figure 3b, which can be attributed to the superior $SiO_2$ quality. However, when extrapolated to the normal operating conditions, the transferred $MoS_2$ sample shows worse performance than the directly-grown $MoS_2$, as discussed in detail below.

The comparison of the device lifetime with different materials and from different processes is shown in Figure 4d. With the presence of transferred h-BN, transferred $MoS_2$, and directly-grown $MoS_2$, the device lifetime at low E-fields can be enhanced from $\sim 10^5$ s (without barrier) to $7.5 \times 10^6$ s, $3.1 \times 10^6$, and $3.7 \times 10^8$ s, respectively, based on the most conservative expectation from the E-model. Thus, we conclude that directly-grown $MoS_2$ gives the best performance in mitigating Cu ion diffusion. Interestingly, devices with transferred h-BN and transferred $MoS_2$ both show similar E-field dependent behaviors, indicating the lifetime of these devices is limited by the film transfer process instead of individual material properties. Defects, cracks and impurities introduced by the mechanical transfer process limit the barrier quality to a large extent. Although optimization of the transfer methods can certainly bring improvement [29], it will remain challenging to realize large-scale transfer of 2D barrier materials with consistent reliability in VLSI technology. We therefore conclude that, directly-grown 2D materials are highly preferred to improve the reliability and device lifetime. It is acknowledged that low temperature growth processes need to be developed to meet the BEOL requirements and prevent thermal damage to the dielectrics. The summary of the material information and the lifetime improvement is listed in Table 1.

**STEM/EDS/EELS analysis.** Besides electrical measurements, STEM in conjunction with EDS and EELS were used for structural analysis and compositional/chemical mapping of the interface and inter-diffusion processes. Devices without a barrier, with transferred h-BN barrier, and with directly- grown $MoS_2$ were analyzed. Each device had an Al cap on top to prevent Cu oxidation and was electrically stressed at 6 MV/cm for 250 s. Under this stress condition, only the control device without a barrier broke down, whereas devices with 2D barriers maintained their initial current values and no breakdown was observed. Figure 5a shows the HAADF (high-angle annular dark-field) STEM cross-sectional image of the $MoS_2$ sample. As the heaviest element, Cu gives the brightest contrast while $SiO_2$/Si and Al appear relatively darker, as expected. Between Al and Cu, there appears to be a uniform layer with a light contrast. EDS suggests that this layer was formed by intermixing of Al with diffused Cu. In both STEM image and EDS line scans, the $Cu/SiO_2$ interface appears sharp with a $MoS_2$ layer clearly detected in between, and Cu diffusion into $SiO_2$ is greatly suppressed.

In the device with the transferred h-BN barrier, the Al and Cu regions were hardly distinguishable, as observed from both the STEM and EDS line scan profile in Figure 5b. This strong inter-diffusion of Al and Cu could be a result of poor Cu adhesion on h-BN. Many pinholes and cracks were observed in Cu deposited on h-BN while rather continuous and uniform Cu formed on $MoS_2$ surface, clearly shown in the scanning electron microscope (SEM) and atomic-force microscopy (AFM) images in Supplementary Figure S2. At the $Cu/SiO_2$ interface, a very weak N signal (not shown) can be identified for the h-BN layers, while the B signal is too small to be detected in EDS. To further verify the existence and position of the h-BN layer, EELS was conducted given its superior resolution for lighter elements. As shown in Supplementary Figure S3, B and N signals were detected between the Cu and $SiO_2$ layer. Similar to the $MoS_2$ sample, Cu diffusion into $SiO_2$ is prevented.

In contrast, the MOS capacitor structure of the control device without any barrier was severely altered by the electrical stress. Figures 5c and 5d are two examples. In Figure 5c, a ball-like feature displaying strong Cu signals was formed. In Figure 5d, a large amount of Cu diffused through $SiO_2$ and reached the Si substrate. This phenomenon has been observed and identified as copper silicide formation by others[30-32], where Cu ions reacted with Si after the diffusion. Note that the devices in previous reports were thermally stressed; while electrical field stress was used in this work, with all measurements at room temperature. This can explain why crystalline copper silicide was not clearly observed here, possibly due to the lack of thermal energy.

Comparing the TEM cross-sections in Figures 5a and 5b, we conclude that Cu started to diffuse into $SiO_2$ in the transferred h-BN device even though no breakdown was measured and the device structure was not changed; whereas no such diffusion was observed at all in the device with directly-grown $MoS_2$ barrier. Similar results were observed in STEM cross-sections of six other positions (three for h-BN and three for $MoS_2$). Therefore, we conclude that directly-grown $MoS_2$ performs better as a diffusion barrier, which is consistent with the TDDB results.

**Conclusion**

The diffusion barrier properties of two types of 2D materials, h-BN and $MoS_2$ have been evaluated using TDDB measurements and by STEM, EDS, and EELS analysis. Predictions of substantial device lifetime improvement are made by analytical models based on experimentally measured times-to-breakdown. For the first time, our work provides strong evidence that these atomically thin 2D materials are capable of suppressing Cu ion diffusion into surrounding dielectrics, identifying them as potential sub-nanometer thin barrier solutions for interconnect technology. We further conclude that direct growth of 2D barriers on dielectric substrates is favored over that of transferred 2D barriers, at least with the present state of the art in both processes. Future studies must focus on a more detailed understanding of the diffusion and breakdown mechanisms through 2D materials, and an optimization of the 2D material deposition to be BEOL compatible.

## Methods

**Preparation of 2D materials**. Cu foils with h-BN grown on both sides by CVD were first coated with polymethyl methacrylate (PMMA) on one side. The side with/without PMMA is identified as the top/bottom side throughout the following descriptions. h-BN on the bottom side was completely etched by Ar plasma. Then, the sample was placed in 1 M iron chloride ($FeCl_3$) solution, with the bottom side facing down, to etch away the exposed Cu. After Cu was completely etched, the sample was immersed in DI water for 10 minutes, followed by 1 M HCl solution for 10 minutes, and another 10 minutes in DI water. The PMMA/h-BN film was then picked up with a 20 nm $SiO_2$ on Si substrate and PMMA was finally dissolved by acetone. $MoS_2$ films were directly grown on 30 nm $SiO_2$ on Si substrates by CVD. Details of the CVD growth can be found elsewhere.[12] To transfer the $MoS_2$ film off the growth substrate, the sample was spin-coated with PMMA and immersed in DI water. A diamond scribe was used to create some scratches at the edges, which allows water to penetrate into the interface of the $MoS_2$ film and the substrate. The PMMA/$MoS_2$ was then detached from the substrate in DI water and transferred to the target substrate. Finally, PMMA was dissolved by acetone.

**Fabrication of MOS capacitor structure**. Heavily-doped Si (resistivity < 5 m$\Omega$-cm) substrates with 20 nm or 30 nm $SiO_2$ were used for the MOS capacitor sample fabrication. After transferring or growing a 2D film, Cu/Al (~30 nm/20 nm) electrodes with diameters of 100 μm were deposited using e-beam evaporation through a shadow mask, with Cu in contact with the 2D material and Al on top. The sample was then coated with photoresist and placed into 6:1 buffered oxide etch (BOE) to etch away the $SiO_2$ on the bottom side of the substrate, followed by 50 nm Al deposition to form an ohmic contact to the Si substrate bottom. Finally, the top photoresist was removed by acetone.

**TEM/EDS/EELS analysis**. STEM cross-sectional samples were prepared with a FEI Nova 200 dual-beam FIB/SEM by using the lift-out method. The region of interest above the Al metal pad was protected during the focused ion beam milling, by depositing $SiO_2$ and Pt layers on top of the sample. Both high resolution transmission electron microscopy (HREM) images, atomic STEM HAADF and bright field (BF) images were obtained in a JEOL ARM200F microscope equipped with a spherical aberration (Cs) corrector (CEOS GmbH, Heidelberg, Germany) and operated at 200 kV. The corrector was carefully tuned by the Zemlin-tableau method with Cs = 0.5 μm and the resolution was demonstrated to be around 1 Å. EDS was performed with an Aztec Energy Advanced Microanalysis System with X-MaxN 100N TLE Windowless 100 mm$^2$ analytical silicon drift detector. Line scan profiles were obtained by scanning the electron probe perpendicularly to the interface of interest. EELS was also performed by using a Gatan parallel electron energy loss spectrometer with better than 1 eV energy resolution.

## Acknowledgments

This work is sponsored by LEAST and SONIC, two of six STARnet Centers, a Semiconductor Research Corporation (SRC) program sponsored by MARCO and DARPA, and is also supported by SWAN Center, a SRC center sponsored by the Nanoelectronics Research Initiative and NIST. This work is also supported in part by the National Science Foundation (NSF) grant CCF-1619062, NSF EFRI 2-DARE grant 1542883, Air Force grant FA9550-14-1-0251, and the Stanford SystemX Alliance. KKHS acknowledges support by the NSF GRFP under Grant No. DGE-114747 and Stanford Graduate Fellowships.

**Table 1 | Material information and lifetime improvement in samples with different barriers.**

| Material | Layer number | Thickness | Lifetime improvement at 0.5 MV/cm (E-model) |
|---|---|---|---|
| Transferred h-BN | 3 - 4 | ~1 – 1.3 nm | ~50× |
| Transferred $MoS_2$ | 1 - 2 | ~0.6 – 1.3 nm | ~20× |
| Directly-grown $MoS_2$ | 1 - 2 | ~0.6 – 1.3 nm | ~1000× |

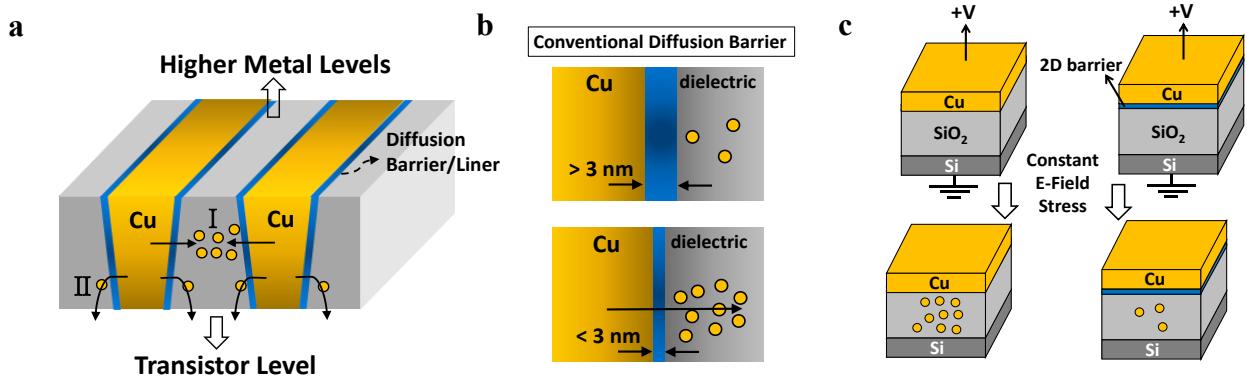

**Figure 1 | Cu diffusion in damascene and test structures for the evaluation of Cu ion diffusion.** (**a**) Schematic of possible Cu diffusion paths in a standard damascene structure: (I) between two neighboring interconnects and (II) to the transistors underneath. (**b**) Conventional materials as diffusion barriers lose their Cu blocking capability when their thicknesses are scaled below ~3 nm. (**c**) MOS capacitors used for barrier property evaluation. Cu ion diffusion into $SiO_2$ with and without 2D barriers under constant E-field stress is illustrated. Note that an Al layer (not shown) was deposited both above Cu and beneath Si. Details of the test structure can be found in "Methods".

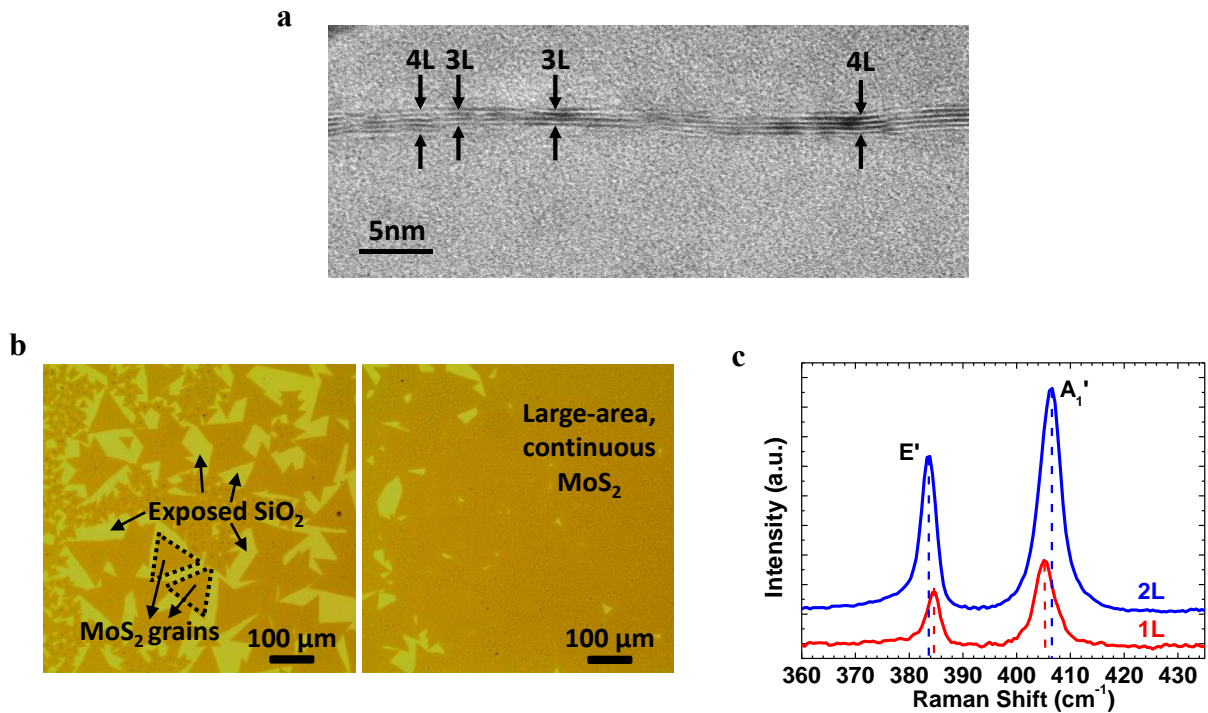

**Figure 2 | Structural and optical characterization of 2D layered materials.** (**a**) STEM cross-sectional image of transferred h-BN. (**b**) Optical images of directly-grown CVD $MoS_2$ films on $SiO_2$. Large $MoS_2$ grains are triangular in shape. The film consists of mostly 1L $MoS_2$ with some 2L regions. The empty areas are exposed $SiO_2$. All measurements were conducted in regions with continuous film coverage. (**c**) Raman spectra of $MoS_2$ on 30 nm $SiO_2$ on Si substrate. Characteristic peaks of 1L and 2L $MoS_2$ can both be identified, with 1L being dominant. The wavelength of the laser used for Raman measurements was 532 nm.

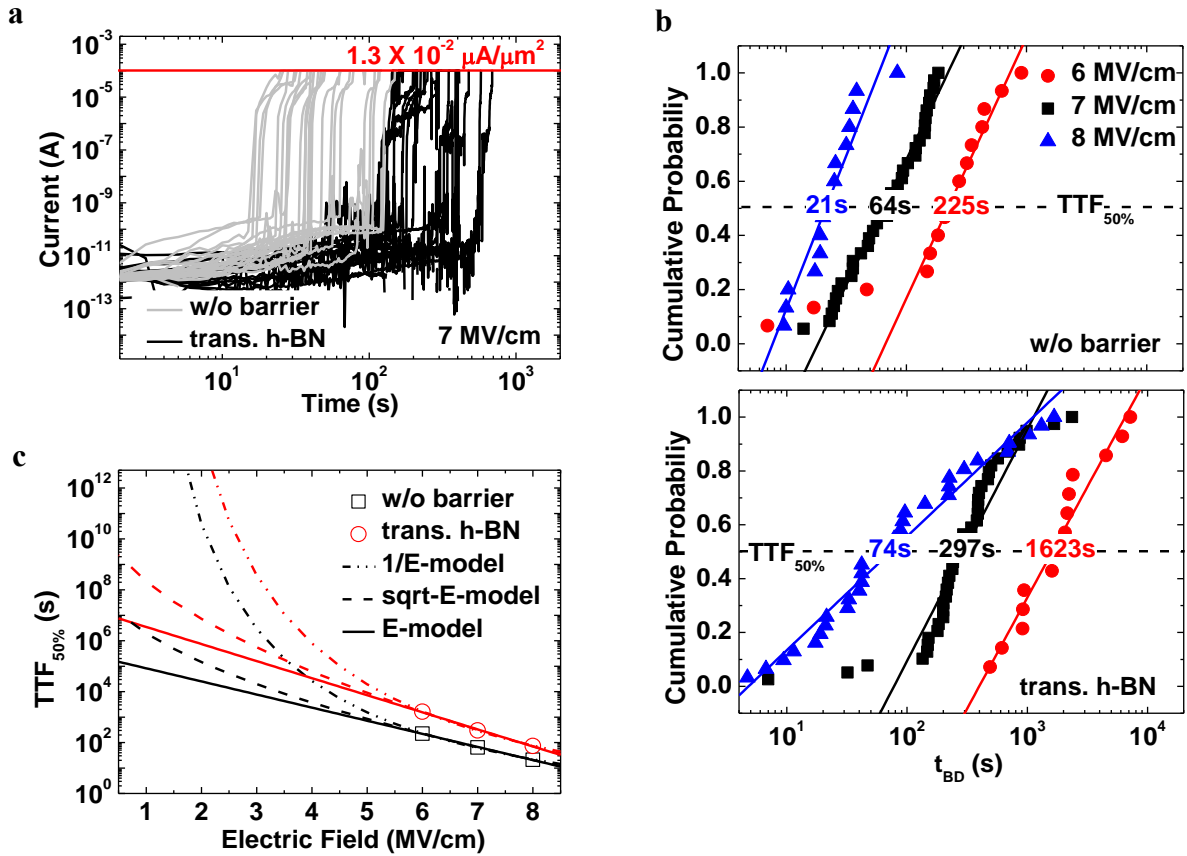

**Figure 3 | Barrier properties of transferred h-BN.** (**a**) Current evolution with time for multiple devices with and without h-BN under the stress condition of 7 MV/cm. Devices without barriers break down earlier in general. (**b**) TDDB results at various E-fields for devices with and without the h-BN barrier. $t_{BD}$ of the h-BN devices is significantly increased. (**c**) Lifetime predictions based on three analytical models. With the presence of h-BN, device lifetime at low fields can be enhanced from $10^5$ s to $7.5 \times 10^6$ s, based on the E-model.

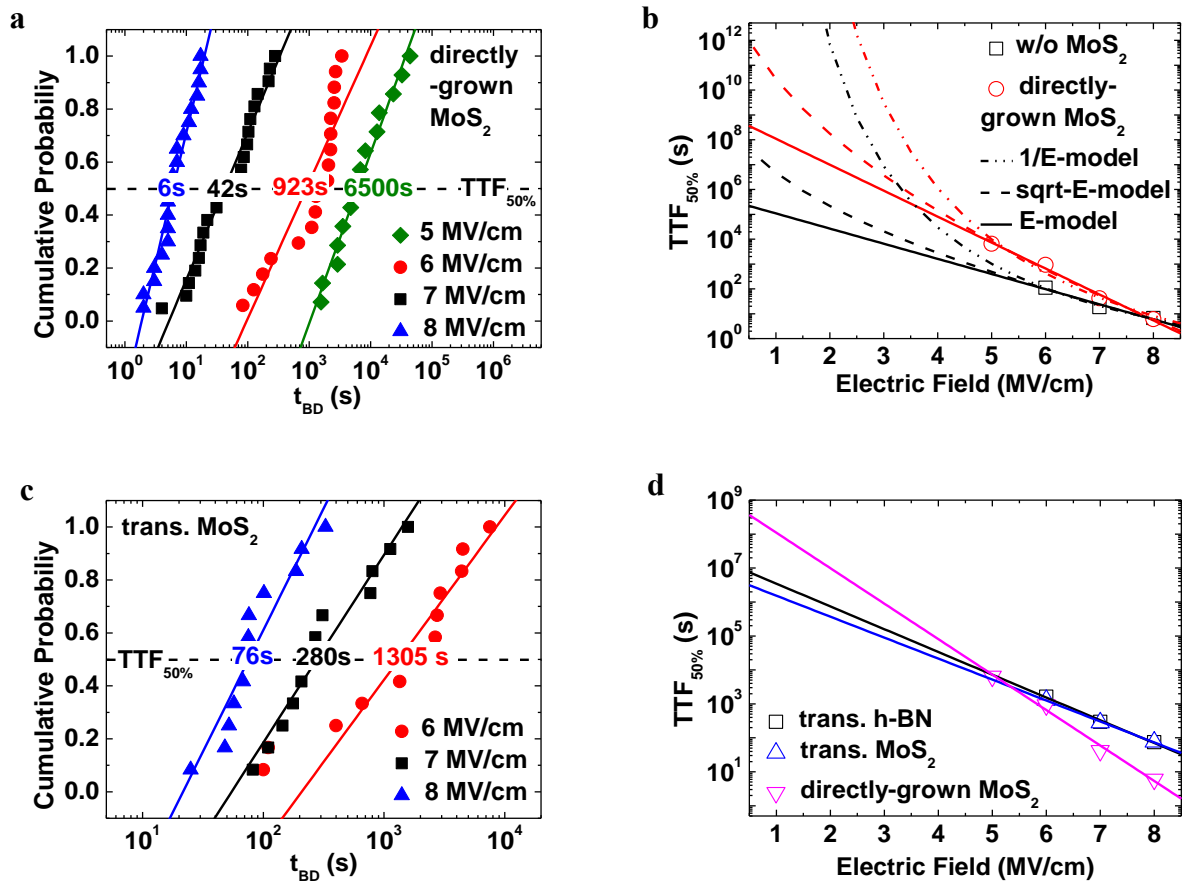

**Figure 4 | Barrier properties of MoS₂ and comparison with the h-BN barrier.** (**a**) TDDB results at various E-fields of devices with directly-grown $MoS_2$ as the diffusion barrier. (**b**) Lifetime prediction of directly-grown $MoS_2$, compared to that of the control sample using various models. With the presence of $MoS_2$, device lifetime can be enhanced from $10^5$ s to $3.7 \times 10^8$ s, based on the E-model. (**c**) Field dependent TDDB results of devices with transferred $MoS_2$. $t_{BD}$ at these high E-fields is enhanced, compared to that of the directly-grown $MoS_2$ devices since the thermal damage of $SiO_2$ was avoided. (**d**) Comparison of the predicted lifetime for devices with different 2D barriers and from different preparation processes, based on the E-model. Directly-grown $MoS_2$ performs the best as a diffusion barrier.

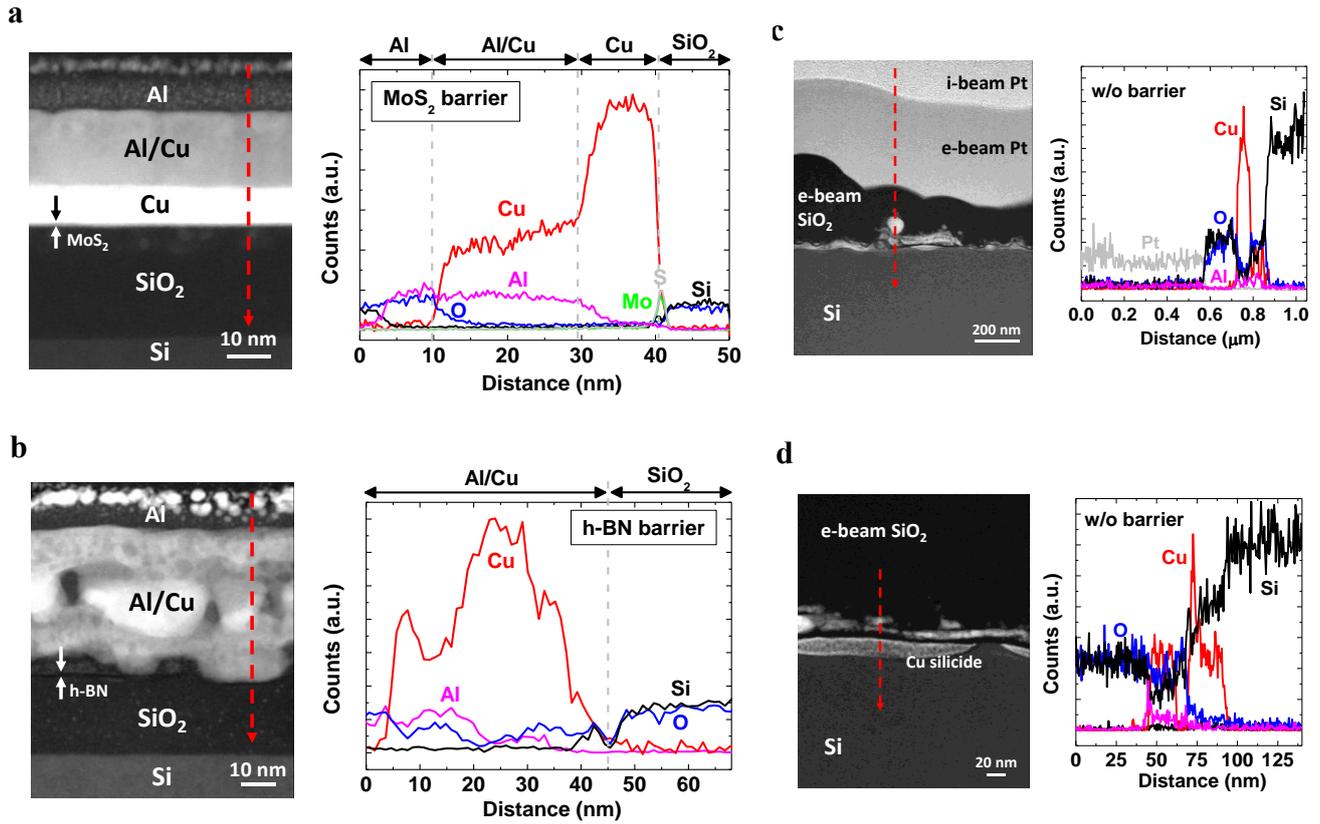

**Figure 5 | Structural, compositional, and chemical analyses.** STEM cross-sections and EDS line scan profiles of devices (**a**) with directly-grown $MoS_2$, (**b**) with transferred h-BN, and (**c-d**) without any barriers. The structures of control devices without barriers were completely damaged after the electrical stress (6 MV/cm; 250 s). The device with either h-BN or $MoS_2$ barrier remained unaltered and Cu signals were barely found in the $SiO_2$ region.

# Supplementary Figures

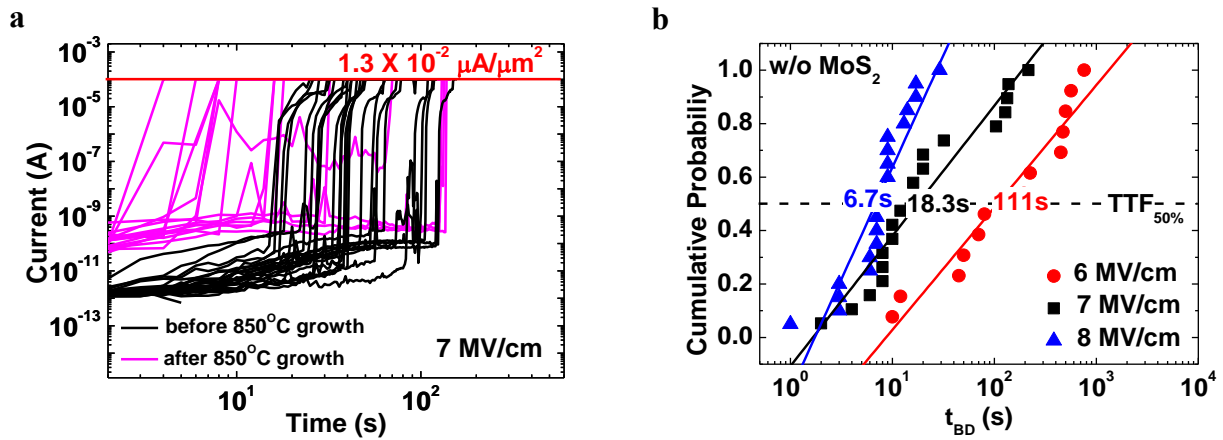

**Supplementary Figure S1 | Effects of thermal stress from high temperature CVD growth.** (**a**) Current evolution with time of multiple devices before and after the thermal stress from the CVD growth. The sulfur-thermal annealed devices (labeled as "after 850 °C growth") went through the same CVD process but intentionally received no $MoS_2$ growth. These devices had higher leakage currents and shorter breakdown time. (**b**) Field dependent TDDB results for the sulfur-thermal annealed devices, which are used as control devices to compare with the devices with directly-grown $MoS_2$ barriers.

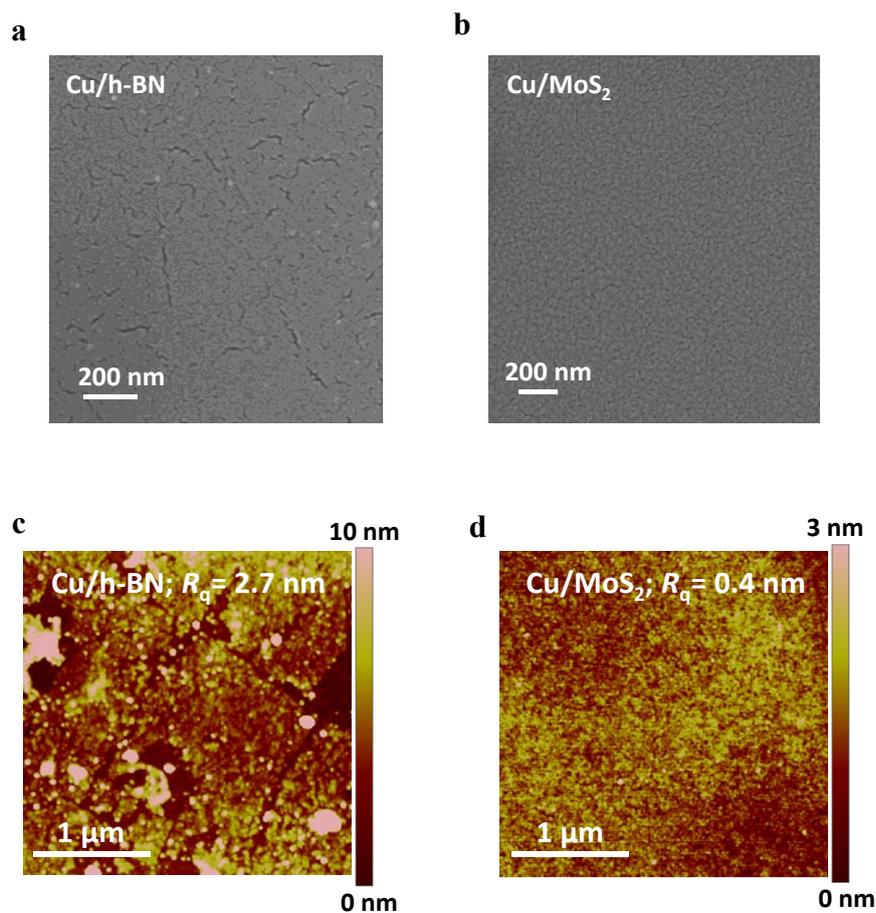

**Supplementary Figure S2 | Cu adhesion on 2D materials.** SEM images of 15 nm Cu surface morphology on (a) h-BN and on (b) $MoS_2$. AFM images of the same Cu surface on (c) h-BN and (d) $MoS_2$. Root-mean-square roughness ($R_q$) of Cu on h-BN and $MoS_2$ are 2.7 nm and 0.4 nm, respectively. Less cracks and roughness are found when Cu was deposited on $MoS_2$, indicating better adhesion.

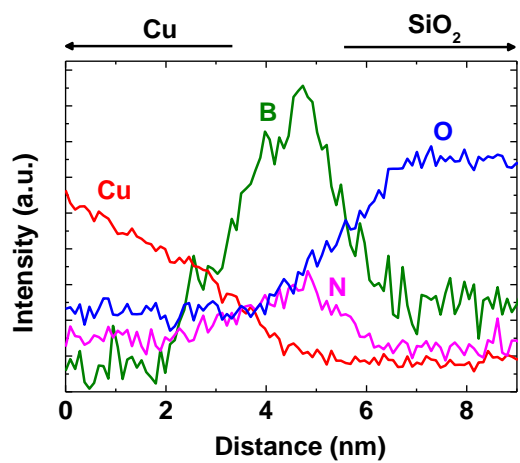

**Supplementary Figure S3 | EELS line scan profile of the device with h-BN barrier.** B and N signals can be detected in-between Cu and SiO₂ layers. The Cu diffusion into SiO₂ was suppressed by h-BN barrier.